\documentclass[floatfix,aps,prb,superscriptaddress,twocolumn]{revtex4-2}

\usepackage[T1]{fontenc}
\usepackage{lipsum}  
\usepackage{graphicx}
\usepackage[english]{babel}
\usepackage{upgreek}
\usepackage{amsmath}
\usepackage{amssymb}
\usepackage{tabls}
\usepackage{dsfont}
\usepackage[colorlinks=true,citecolor=blue,linkcolor=red,urlcolor=blue]{hyperref}
\usepackage{bookmark}
\usepackage{physics}

\makeatletter
\def\bbl@set@language#1{%
	\edef\languagename{%
		\ifnum\escapechar=\expandafter`\string#1\@empty
		\else\string#1\@empty\fi}%
	\@ifundefined{babel@language@alias@\languagename}{}{%
		\edef\languagename{\@nameuse{babel@language@alias@\languagename}}%
	}%
	\select@language{\languagename}%
	\expandafter\ifx\csname date\languagename\endcsname\relax\else
	\if@filesw
	\protected@write\@auxout{}{\string\select@language{\languagename}}%
	\bbl@for\bbl@tempa\BabelContentsFiles{%
		\addtocontents{\bbl@tempa}{\xstring\select@language{\languagename}}}%
	\bbl@usehooks{write}{}%
	\fi
	\fi}
\newcommand{\DeclareLanguageAlias}[2]{%
	\global\@namedef{babel@language@alias@#1}{#2}%
}
\makeatother

\newcommand\varpm{\mathbin{\vcenter{\hbox{%
  \oalign{\hfil$\scriptstyle+$\hfil\cr
          \noalign{\kern-.3ex}
          $\scriptscriptstyle({-})$\cr}%
}}}}
\newcommand\varmp{\mathbin{\vcenter{\hbox{%
  \oalign{$\scriptstyle({+})$\cr
          \noalign{\kern-.3ex}
          \hfil$\scriptscriptstyle-$\hfil\cr}%
}}}}

\DeclareLanguageAlias{en}{english}

\begin{document}

\title{Phonon-induced frequency shift in semiconductor spin qubits}

\author{Irina Heinz}
\email{i.heinz@fz-juelich.de}
\affiliation{Department of Physics, University of Konstanz, D-78457 Konstanz, Germany}
\affiliation{Peter Grünberg Institute, Theoretical Nanoelectronics, Forschungszentrum Jülich, D-52425 Jülich, Germany}
\affiliation{Institute for Quantum Information, RWTH Aachen University, D-52056 Aachen, Germany}
\author{Jeroen Danon}
\affiliation{Department of Physics, Norwegian University of Science and Technology, NO-7491 Trondheim, Norway}
\author{Guido Burkard}
\email{guido.burkard@uni-konstanz.de}
\affiliation{Department of Physics, University of Konstanz, D-78457 Konstanz, Germany}


\begin{abstract}
    Spin qubits have proven to be a feasible candidate for quantum computation, and some realizations of spin qubits already benefit from advanced device manufacturing in the semiconductor industry. Compared to superconducting platforms, spin qubits can operate at higher temperatures from tens of millikelvin up to a few kelvin. However, recent experiments \cite{PhysRevApplied.19.044078, PhysRevX.13.041015} show a non-trivial and often non-monotonic dependence of the spin qubit frequency on the temperature, featuring a region of decreased sensitivity to temperature fluctuations. In this work, we aim to gain insight into the physics behind such temperature shifts in the low-temperature limit.  Investigating the spin qubits' interaction with phonon modes of the host material, we can explain some of the key features of the observed behavior and estimate the temperature sweet spot for the qubit frequency shift.
\end{abstract}


\maketitle

\section{Introduction}

In the last decades, various physical implementations for quantum computation have been developed, including superconducting circuits, ion traps, and spin qubits~\cite{deleon2021}. The latter are considered a promising candidate for a scalable quantum computing platform as they benefit from advanced fabrication techniques of the established semiconductor industry, which in principle allow for the manufacturing of a large number of qubits on a small chip~\cite{Ha_2021, Künne2024, koch2024industrial300mmwaferprocessed, George_2024, Steinacker_2025}. A spin qubit can be hosted in a quantum dot formed in a semiconductor heterostructure through voltages applied to gate electrodes on top, which is filled with an electron or hole. The spin of the respective charge carrier is then used as the computational unit for quantum computation \cite{RevModPhys.95.025003}.

Compared to superconducting platforms, spin qubits can operate at higher temperatures, from tens of millikelvin to a few kelvin~\cite{Mills_2022, noiri2021fast, PhysRevX.13.041015, Camenzind2022, Huang_2024}.  However, recent experiments~\cite{PhysRevApplied.19.044078, PhysRevX.13.041015} show a non-trivial dependence of the spin-qubit frequency on the temperature, indicating a region of decreased sensitivity to temperature fluctuations, which is not fully understood. Microwave control of spin qubits not only enables quantum operations but also causes additional heating, making it important to understand how an increased temperature influences the qubit~\cite{PhysRevApplied.19.044078, PhysRevX.13.041015, Freer_2017, Takeda_2018, Hendrickx_2020, Zwerver_2022, Philips_2022, doi:10.1126/sciadv.add9408, ye2025measuringpulseheatingsi}. 

Although two-level fluctuators in the interacting random dipole defect model of Ref.~\cite{PhysRevResearch.6.013168} could in principle explain some of the observations in the experiment, in this model the correlations between the positions of the two-level fluctuators and the strength and direction of their random fields that are required to fit the model to the experiment are not well understood. Furthermore, results from Ref.~\cite{sato2024simulationtemperaturedependentquantumgates} show that a temperature-dependent charge-noise model is not sufficient to describe the experimentally observed shift due to microwave pulse heating. However, combining it with an experimental fit based on the Debye model quantitatively reproduces the experimental data which exhibits a megahertz shift. These results indicate a connection between thermal activation and a temperature-dependent qubit frequency shift and raises the question of whether phonons that couple to the electron can have a noticeable effect on the spin splitting.

\begin{figure}[b]
	\centering
	\includegraphics[width=0.48\textwidth]{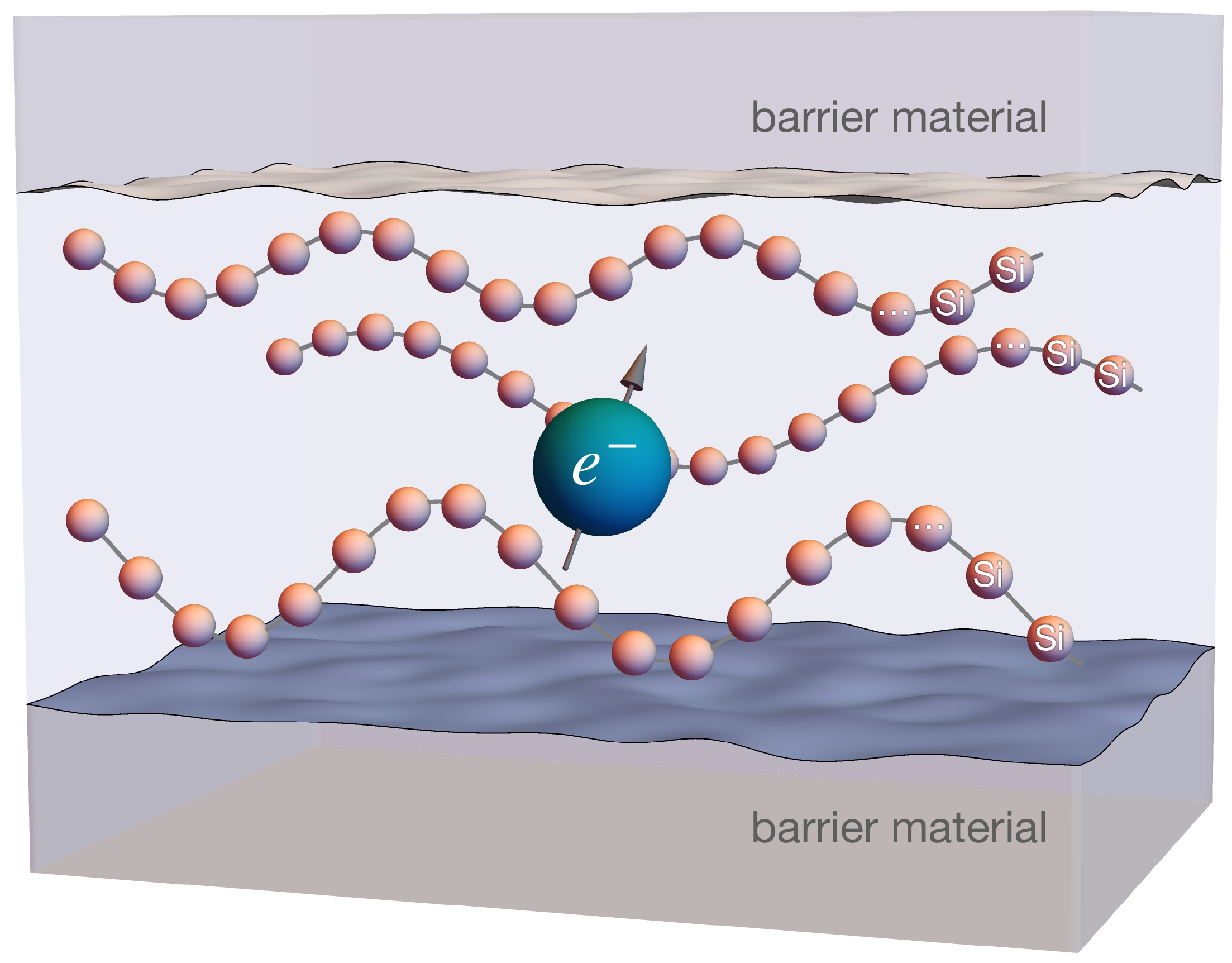}
	\caption{Schematic of an electron spin qubit in a silicon quantum dot with thermally activated acoustic waves in the material.}
	\label{Fig:Schematic}
\end{figure}

The interaction of a spin qubit with phonons was extensively studied in the context of qubit relaxation \cite{PhysRevB.64.125316, PhysRevLett.93.016601, hu2011two,PhysRevB.89.075302, li2020hole, PhysRevB.104.085309, brooks2024phonon,PhysRev.118.1523, PhysRev.101.944}. Here, we focus on the coherent contributions to the Hamiltonian due to phonons that can give rise to shifts in the eigenenergies.
We investigate the electron-phonon interaction in more detail and provide insight into the role of phonons for temperature-dependent frequency shifts. We consider contributing phonons at different energy scales provided by the Zeeman, valley, and orbital splittings (Fig.~\ref{Fig:Model}a). We derive an approximate frequency shift due to the interaction with single phonon modes and extend our findings to multiple modes. Our results reproduce some of the observed features; however, the order of magnitude of the effective shift depends on various parameters, which at this stage we cannot fully describe.

This paper is organized as follows. First, starting from a description of the entire quantum dot, we derive an effective low-energy Hamiltonian in which the spin qubit effectively couples to the phonons in the solid. Next, we outline the general treatment of the three relevant energy scales involved, where we start with a single phonon mode and afterwards include the thermal statistics of all acoustic phonons. Finally, we derive the energy shifts at the three relevant energy scales and calculate the temperature dependence numerically.

\section{Theoretical model}
\begin{figure}
	\centering
	\includegraphics[width=0.45\textwidth]{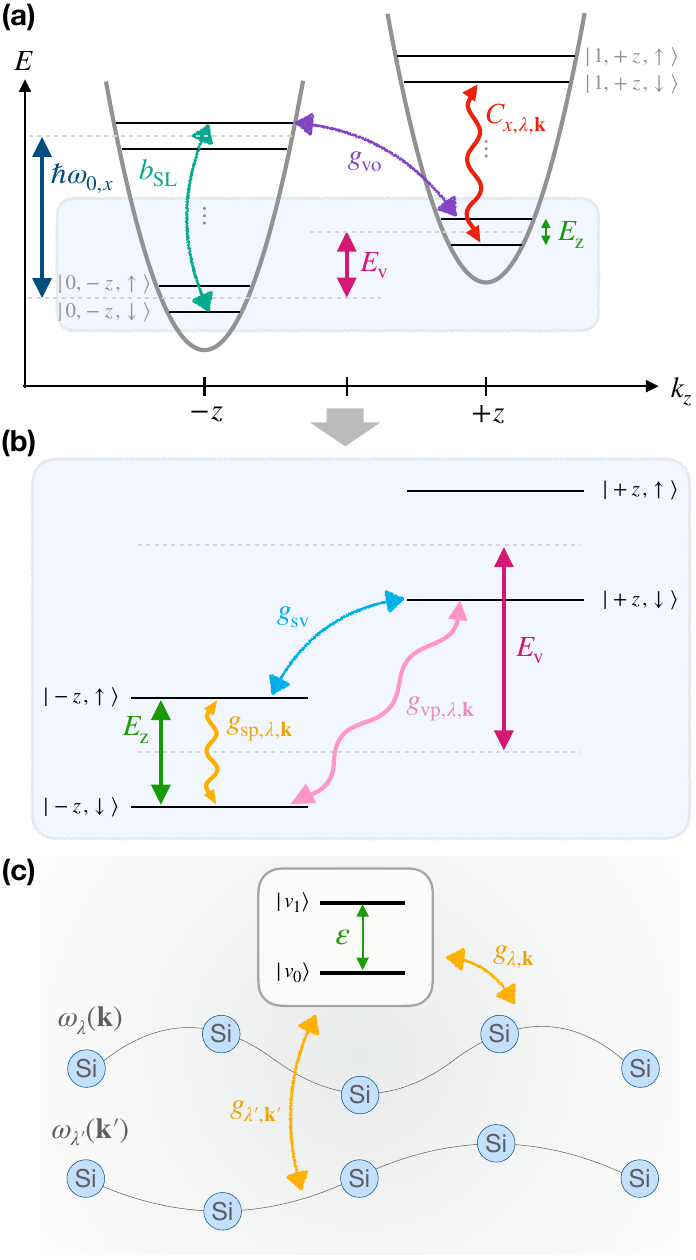}
	\caption{(a) Schematic of the energy scales relevant for the electron confined in a Si quantum dot. The orbital energy splitting $\hbar \omega_{0,x}$ is determined by the harmonic confinement potential (grey) in $x$-direction. The two low-lying conduction band minima of the silicon heterostructure are located at $\pm z$ in $\mathbf{k}$-space, yielding two valleys split by energy $E_{v}$. A magnetic field splits the spin states by $E_{z}$. A micromagnet induces a slanting magnetic field at the dot position and gives rise to an effective spin-orbit interaction with strength~$b_{\rm SL}$. Imperfections at the heterostructure interface can lead to a valley-orbit interaction~$g_{\rm vo}$. Finally, the electron-phonon interaction $C_{x,\lambda, \mathbf{k}}$ induces transitions between the orbital levels under phonon emission or absorption. The light-blue box indicates the low-energy subspace relevant for the spin qubit. (b)~Results of the Schrieffer-Wolff transformation to derive an effective Hamiltonian including the spin-phonon coupling $g_{\rm sp}$, valley-phonon coupling $g_{\rm vp}$, and spin-valley coupling $g_{\rm sv}$. (c)~Effective treatment using a two-level system (e.g., $\ket{v_0} = \ket{\downarrow}$ and $\ket{v_1} = \ket{\uparrow}$) split by $\varepsilon$ and its coupling to a phonon bath with strength $g_{\lambda, \mathbf{k}}$. In case of a spin coupling to phonons, this corresponds to $\ket{v_0} = \ket{\downarrow}$ and $\ket{v_1} = \ket{\uparrow}$,  $g_{\lambda, \mathbf{k}} = g_{{\rm sp}, \lambda, \mathbf{k}}$, and $\varepsilon = E_{z}$.}
	\label{Fig:Model}
\end{figure}
To describe a single electron spin in a semiconductor quantum dot in silicon (Fig.~\ref{Fig:Schematic}), we start with the following Hamiltonian,
\begin{align}
    H = H_{\rm QD} + H_{v} + H_{z} + H_{\rm vo} + H_{\rm so} + H_{p} + H_{\rm ep}, \label{Eq:Hall}
\end{align}
which describes various states and processes that are depicted in Fig.~\ref{Fig:Model}a. In the figure, energy splittings are indicated by straight double arrows, the couplings between energy levels are depicted as curved double arrows and phonon-induced transitions are represented by wavy double arrows. The Hamiltonian contains the quantum dot confinement potential
\begin{align}
    H_{\rm QD} = \sum_{i \in \{x,y,z\}} \hbar \omega_{0,i} a_{i}^{\dagger} a_i,
\end{align}
 where $a_i$ and $a_i^{\dagger}$ are the orbital ladder operators.
For simplicity, we assume a harmonic potential in all three dimensions. In reality, the confinement in the $z$-direction will be asymmetric as the electron wave function is pushed towards the upper interface, typically Si/SiGe or Si/SiO$_2$, by the applied gate voltages. However, for the purpose of understanding the phonon interaction with the system, harmonic confinement shall be a valid approximation. 
The second and third terms correspond to the valley and Zeeman splitting in the valley and spin eigenbasis, respectively,
\begin{align}
    H_{v} = -\frac{E_{v}}{2} \tau_z, \hspace{0.5cm}
    H_{z} = -\frac{E_{z}}{2} \sigma_z,
\end{align}
where $\tau_i$ and $\sigma_i$, with $i\in \{x, y, z\}$, are the valley pseudospin and electron spin Pauli operators, respectively.
Inhomogeneity of the interface can lead to a valley-orbit coupling $g_{\rm vo}$ in Si/SiGe~\cite{PhysRevB.81.115324, PhysRevB.73.235334, Lima_2023} or Si/SiO$_2$~\cite{Rahman2011}, that we simplify as
\begin{align}
    H_{\rm vo} = g_{\rm vo} (a_x^{\dagger} + a_x) \tau_x,
\end{align}
where we have left out couplings in other spatial directions, which could enter as additional terms in the overall shift, but do not add a qualitatively different ingredient.
The spin-orbit coupling mechanism for electrons in silicon is rather weak, which is why we neglect it at this stage. Instead, we take into account a magnetic gradient field 
\begin{align}
    H_{\rm so} = b_{\rm SL} (a_x^{\dagger} + a_x) \sigma_x
\end{align}
induced by a micromagnet that is typically employed to enable electrical control of the spin via electric-dipole spin resonance (EDSR). Here $b_{\rm SL}$ is the slanting magnetic field in energy units. A similar expression can be derived in materials with strong intrinsic spin-orbit mixing \cite{PhysRevB.74.165319, PhysRevB.64.125316, PhysRevLett.93.016601}.

The phonons are taken into account as a sum over branches $\lambda$ and wave vectors $\mathbf{k}=k(\sin(\theta) \cos(\phi), \sin(\theta)\sin(\phi), \cos(\theta))$ via
\begin{align}
    H_{p} = \sum_{\lambda, \mathbf{k}} \hbar \omega_{\lambda}(\mathbf{k}) b_{\lambda,\mathbf{k}}^{\dagger} b_{\lambda,\mathbf{k}},
\end{align}
where $b_{\lambda,\mathbf{k}}$ and $b_{\lambda,\mathbf{k}}^{\dagger}$ are the respective phonon annihilation and creation operators. We can approximate a linear dispersion $\omega_{\lambda}(\mathbf{k}) \approx v_{\lambda} k$ for low-energy acoustic phonons in the temperature regime of interest, where $v_{\lambda}$ is the sound velocity in branch $\lambda$. In the presence of lattice vibrations, the confined electron is exposed to deformations in the potential, effectively interacting with the phonons via \cite{kittel2005introduction, yu2010fundamentals, PhysRev.118.1523, PhysRev.101.944}
\begin{align}\label{eq:e-ph}
    H_{\rm ep} = \sum_{\lambda, \bf{k}} \sqrt{\frac{\hbar k}{2 \rho_{\rm Si} V v_{\lambda}}} \Xi_{\lambda,\bf{k}}  \left( b_{\lambda,\bf{k}} e^{i\bf{kr}} - b_{\lambda,\bf{k}}^{\dagger} e^{-i\bf{kr}}  \right) , 
\end{align}
where $\rho_{\rm Si}$ is the mass density of silicon, $V$ is the volume of the crystal, and $\Xi_{\lambda,\bf{k}}$ are the branch-dependent and $\mathbf{k}$-dependent deformation potential strengths. The deformation potentials 
can be calculated from the strain tensor and the phonon polarization vector resulting in $\Xi_{l,\bf{k}} = \Xi_{d} + \Xi_{u} \cos^2(\theta)$ and $\Xi_{t,\bf{k}} = \Xi_{u} \sin(\theta) \cos(\theta)$ for the longitudinal and transverse branch with $\Xi_{u} = 8.77$~eV and $\Xi_{d} = 5$~eV, for strained silicon in the [001] direction~\cite{PhysRevB.89.075302, PhysRevB.104.085309, PhysRev.101.944}. The transverse and 
longitudinal sound velocities are $v_{t} = 5420$~m/s and 
$v_{l}=9330$~m/s, and the mass density of silicon is $\rho_{\rm Si} = 2330$~kg/m$^3$.

Since we are interested in the low-energy subspace of the full Hamiltonian, we will restrict ourselves to the matrix element that connects the orbital ground state to the first orbitals. For a harmonic confinement potential in all three dimensions, we can write the states of the three-dimensional harmonic oscillator as $F_{n_x, n_y, n_z} = \psi_{n_x}(x) \psi_{n_y}(y) \psi_{n_z}(z)$ using the one-dimensional solutions $\psi_{0}(x) =  e^{-\frac{x^2}{2 \ell_x^2}}/(\pi \ell_x^2)^{1/4}$ and $ \psi_{1}(x) =  \sqrt{2}x e^{-\frac{x^2}{2 \ell_x^2}}/(\pi^{1/4} \ell_x^{3/2} )$ with $\ell_x=\sqrt{\hbar/(m_x \omega_{0,x})}$, and similar for the other dimensions. Here, $\hbar\omega_{0,i}$ are the orbital energies of the quantum dot, and $m_i$ is the corresponding effective mass of the electron in the dot. The transverse and longitudinal effective masses in silicon are $m_t = 0.19 m_e$ and $m_l= 0.92 m_e$, respectively, with electron mass $m_e$, corresponding to in-plane and out-of-plane directions of the heterostructure. We focus on the matrix element for the transition between the ground and one of the first excited states, $\langle F_{000}| H_{\rm ep} |F_{100}\rangle$ and analogously for the excited states $F_{010}$ and $F_{001}$ \cite{PhysRevB.89.075302}, which allows to reduce Eq.~(\ref{eq:e-ph}) to 
\begin{align}
    H_{\rm ep} = \sum_{i \in \{x,y,z\}} \sum_{\lambda, \bf{k}} C_{i,\lambda,\bf{k}}  \left( b_{\lambda,\bf{k}} +  b_{\lambda,\bf{k}}^{\dagger}  \right) (a_i^{\dagger} + a_i), \label{Eq:Hep-1stexcitedstate}
\end{align}
where $C_{i,\lambda,\bf{k}} = - \sqrt{\frac{\hbar k}{\rho_{\rm Si} V v_{\lambda}}} \Xi_{\lambda,\bf{k}} \frac{\ell_i k_i}{2} e^{-\frac{\ell_x^2k_x^2 + \ell_y^2k_y^2 + \ell_z^2k_z^2}{4}}$.
In Fig.~\ref{Fig:PhononCoupling} we plot the longitudinal and transversal couplings $C_{x,l, \mathbf{k}}$ and $C_{x,t, \mathbf{k}}$ as a function of the phonon energy.
We see a $\omega^{3/2}$-behavior at low energies, where the dipole approximation would be appropriate, and a Gaussian decay at energies comparable to the orbital splitting on the quantum dot.

\subsection{Low-energy subspace} \label{Sec:Low-energy-supspace}
Typically the energy scales of spin qubits in Si quantum dots have the hierarchy $E_{z} \ll E_{v} \ll \hbar \omega_{0,i}$. Since the couplings to the first orbital are rather small $|g_{\rm vo}|, |b_{\rm SL}|, |C_{i,\lambda,\bf{k}}| \ll \hbar\omega_{0,i}$ we can perform a Schrieffer-Wolff transformation on the Hamiltonian in Eq.~\eqref{Eq:Hall} to obtain an effective Hamiltonian in the spin-valley space
\begin{align}
    H_{\rm eff} = H_{v} + H_{z} + H_{p} + H_{\rm sv} + H_{\rm vp} + H_{\rm sp}, \label{Eq:Heff-all}
\end{align}
with
\begin{align}
    H_{\rm sv} &= g_{\rm sv} \tau_x \sigma_x, \\ 
    H_{\rm vp} &= \sum_{\lambda, \bf{k}} g_{\rm vp, \lambda,\bf{k}} \left( b_{\lambda,\bf{k}} +  b_{\lambda,\bf{k}}^{\dagger}  \right) \tau_x,\\
    H_{\rm sp} &= \sum_{\lambda, \bf{k}} g_{\rm sp, \lambda,\bf{k}} \left( b_{\lambda,\bf{k}} +  b_{\lambda,\bf{k}}^{\dagger}  \right) \sigma_x. \label{Eq:Hsp}
\end{align}
We have omitted two-phonon terms that do not couple to the spin qubit but arise from the transformation. 
The spin-valley coupling is determined by $g_{\rm sv} = -2 g_{\rm vo} b_{\rm SL}/(\hbar\omega_{0,x})$, the effective valley-phonon coupling is given by $g_{\rm vp,\lambda,\bf{k}} = - 2 g_{\rm vo} C_{x,\lambda,\bf{k}}/(\hbar\omega_{0,x})$, and the effective spin-phonon coupling is $g_{\rm sp,\lambda,\bf{k}} = - 2 b_{\rm SL} C_{x,\lambda,\bf{k}}/(\hbar\omega_{0,x})$. The effective Hamiltonian is illustrated in Fig.~\ref{Fig:Model}b.

\subsection{General treatment} \label{Sec:GeneralTreatment}
In the following, we will restrict our analysis to the low-energy dynamics of the system, considering the lowest two energy levels with level spacing $\varepsilon$, with an interaction strength to phonons given by $g_{\lambda,\bf{k}}$, as depicted in Fig.~\ref{Fig:Model}c. We will assign $\varepsilon$ and $g_{\lambda,\bf{k}}$ to (i) $E_{z}$ and $g_{\rm sp, \lambda,\bf{k}}$ for the low-energy spin qubit subspace, (ii) $E_{v}$ and $g_{\rm vp, \lambda,\bf{k}}$ to capture also effects on the valley splitting and (iii) $\omega_{0,x}$ and $C_{x,\lambda,\bf{k}}$ to estimate effects at the orbital energy scale. The Hamiltonian has the form 
\begin{align}
    \tilde{H} =& -\frac{\varepsilon}{2} \begin{pmatrix} 1&0 \\ 0&-1\end{pmatrix} + H_{p} + \sum_{\lambda, \bf{k}} g_{\lambda,\bf{k}} \left( b_{\lambda,\bf{k}}
    + b_{\lambda,\bf{k}}^{\dagger}  \right) \begin{pmatrix} 0&1 \\ 1& 0\end{pmatrix}. \label{Eq:Hgeneral}
\end{align}
We refer to the basis states of the two-level system as $\ket{v_0}$ and $\ket{v_1}$ and will discuss the meaning of the corresponding states in the following sections.
To understand the effect of phonons on the two-level system, we first consider a single phonon frequency $\omega_{\lambda}(\mathbf{k})= \omega$ with coupling $g_{\lambda, \mathbf{k}} =g$ in Eq.~\eqref{Eq:Hgeneral}. In this case, we obtain the Hamiltonian 
\begin{widetext}
\begin{align}
\vcenter{\hbox{\includegraphics[width=0.88\textwidth]{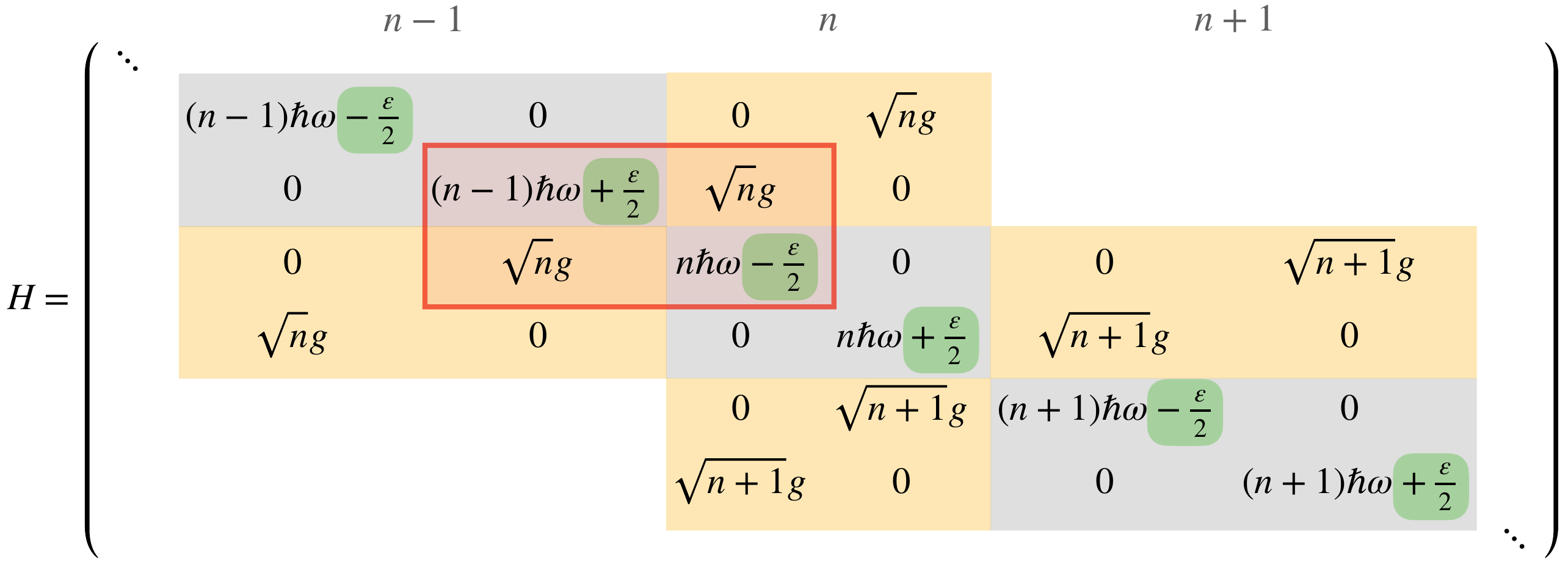}}},\label{Eq:GeneralSingleModeHamiltonian} 
\end{align}
\end{widetext}
where the gray diagonal blocks correspond to different phonon numbers $n$, and each block contains the $\ket{v_0}$ and $\ket{v_1}$ states of the two-level system with energies shown in green. The phonon-mediated coupling terms in Eq.~\eqref{Eq:Hgeneral} are highlighted in yellow. 

\begin{figure}[t]
	\centering
	\includegraphics[width=1\linewidth]{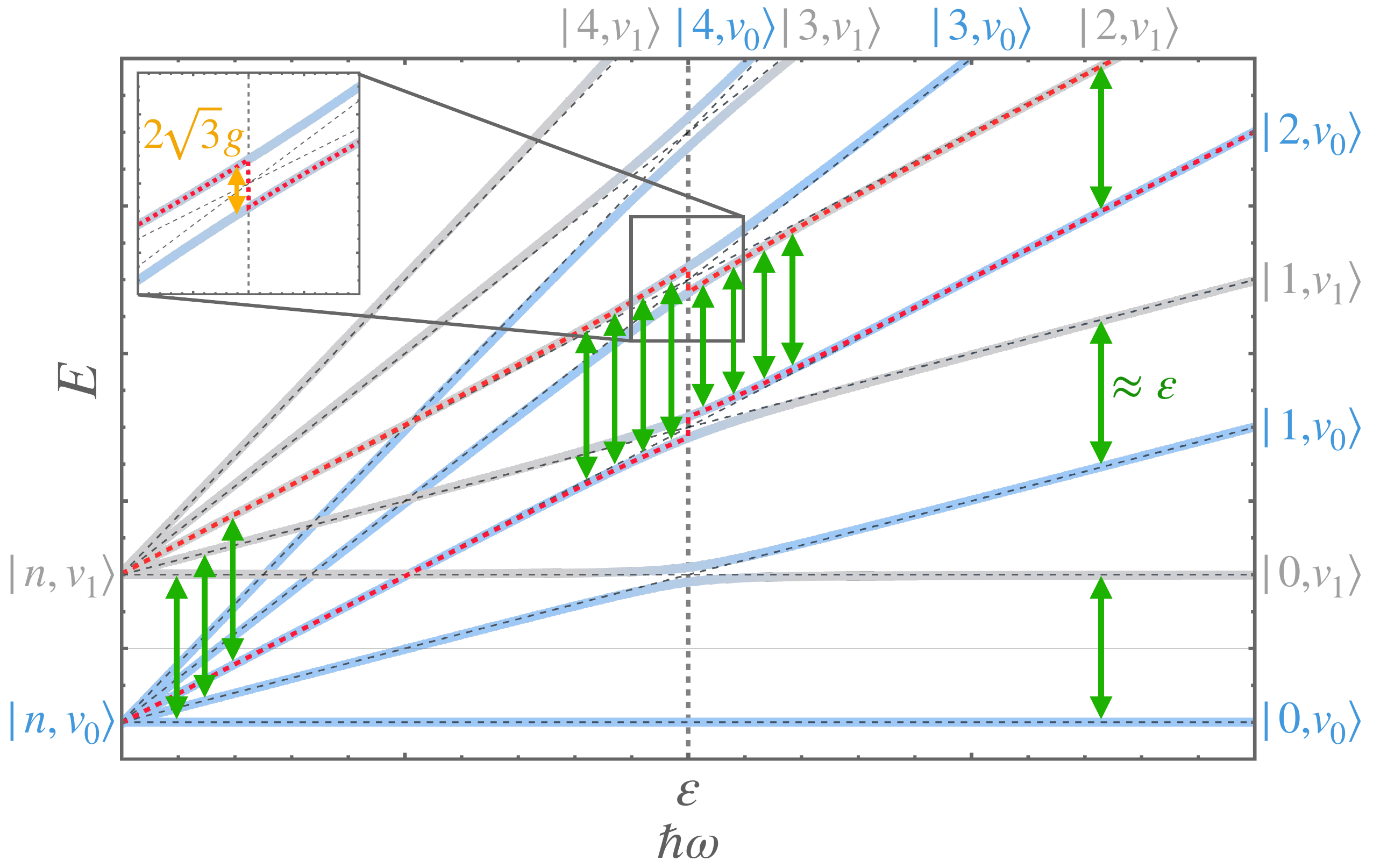}
	\caption{Eigenenergies obtained from diagonalizing the blocks in Eq.~\eqref{Eq:GeneralSingleModeHamiltonian} similar to the block in the red box, as a function of the phonon energy and for different phonon numbers $n$. At $\hbar \omega \approx 0$, the states $\ket{n,v_0}$ and $\ket{n,v_1}$ (depicted in blue and gray) are well separated by approximately the energy~$\varepsilon$ (shown as green double arrows). Without the interaction between spin and phonons, $g=0$, one obtains the black dashed lines with a crossing at the resonance. The presence of an interaction yields an anticrossing of width $2\sqrt{n} g$ at the energy $\hbar \omega = \varepsilon$ leading to a role change of the energy levels. An example for the definition of the two qubit levels with phonon occupation number~$n=2$ is shown as red dashed lines featuring a jump at the resonance. The green double arrows show the resulting abrupt change in frequency at the resonance.}
	\label{Fig:EnergyLevels}
\end{figure}

In a low-temperature and weak-coupling model, we can assume that the number of phonons, determined by the Bose-Einstein statistics, times the coupling constant is small compared to the sum of the level spacing and phonon frequency, and thus $|\sqrt{n} g| \ll |\varepsilon+\hbar\omega|$. We discuss this assumption in the following sections and provide further details in Appendix~\ref{App:PhononsModelValidity}. In this case, we can calculate the hybridized eigenvalues $(n-\frac{1}{2})\hbar\omega \pm  \frac{1}{2}\sqrt{4ng^2+(\varepsilon-\hbar\omega)^2}$ by diagonalizing the red block in Eq.~\eqref{Eq:GeneralSingleModeHamiltonian}. 
The resulting energy levels, depending on the phonon frequency, are depicted as light gray and light blue lines in Fig.~\ref{Fig:EnergyLevels} for different values of $n$, and feature an avoided crossing of width $2\sqrt{n} g$ at resonance $\hbar \omega=\varepsilon$. The black dashed lines show the energies of the non-interacting Hamiltonian, i.e., $g=0$.

In the far off-resonant regime $|\hbar \omega - \varepsilon| \gg g$, the states $\ket{n,v_0}$ and $\ket{n,v_1}$ are well defined, with an energy separation $\approx \varepsilon$. When approaching the resonance from $\hbar \omega <\varepsilon$ ($\hbar \omega > \varepsilon$) these eigenstates develop an increased (decreased) level spacing (see green double arrows for the case of $n=2$). Exactly at resonance, i.e., a phonon energy $\hbar \omega = \varepsilon$, all eigenstates are equal superpositions of $\ket{v_0}$ and $\ket{v_1}$. However, on the left and right side of the resonance, the hybridized states are not equally superposed, and we can assign a qubit state according to the larger contributions of $\ket{v_0}$ or $\ket{v_1}$. This results in the following definition of the energies of $\ket{v_0}$ and $\ket{v_1}$ in the presence of $n$ phonons,
\begin{align}
E_{n-1,n} ^{\pm} = 
\left(n-\frac{1}{2}\right) \hbar \omega \pm \frac{(\varepsilon- \hbar \omega)}{2}  \sqrt{\frac{4ng^2}{(\varepsilon- \hbar \omega)^2}+1} \label{Eq:redblockeigenvalues},
\end{align}
where $\pm$ corresponds to $\ket{v_1}$ and $\ket{v_0}$, respectively.
These energies are indicated for the case of $n=2$ with the red dashed lines in Fig.~\ref{Fig:EnergyLevels}, showing a jump at the resonance $\hbar \omega=\varepsilon$.
The resulting effective level spacing between $\ket{v_1}$ and $\ket{v_0}$ (green double arrows in Fig.~\ref{Fig:EnergyLevels}) follows as 
\begin{widetext}
    \begin{align}
	\tilde{E}_{n} (g) &= E_{n,n+1} ^{+} -E_{n-1,n} ^{-} 
	= \hbar\omega + \frac{(\varepsilon -\hbar\omega)}{2}   \left( \sqrt{\frac{4(n+1)g^2}{(\varepsilon-\hbar\omega)^2}+1} +   \sqrt{\frac{4ng^2}{(\varepsilon-\hbar\omega)^2}+1} \right). \label{Eq:DeltaE}
\end{align}
\end{widetext}
For a far off-resonant phonon energy $|\hbar \omega - \varepsilon|\gg g$ we can approximate $\sqrt{1+x^2} \approx 1+x^2/2$ in Eq.~\eqref{Eq:DeltaE}, yielding
\begin{align}
\tilde{E}_{n} (g) \approx \varepsilon + \frac{g^2}{\varepsilon-\hbar \omega} + \frac{2ng^2}{\varepsilon- \hbar \omega} . \label{Eq:Lamb-shift}
\end{align}
These corrections to the undisturbed level spacing can be interpreted as the Lamb and ac Stark shift. Again, for $\hbar \omega < \varepsilon$,  we obtain a positive frequency shift, and for $\hbar \omega > \varepsilon$, a negative frequency shift. Only the second correction depends on the phonon occupation number and can thus give rise to a temperature dependence of the shift.
Using the upper or lower branch of two anticrossing states, depending on the relative weight of their components, when calculating a qubit energy splitting might seem a bit arbitrary.
In fact, a level splitting in the $\{\ket{v_0},\ket{v_1}\}$ subspace at the resonance is not well-defined.
Close to resonance, the effective qubit dynamics will always involve more than two levels and depending on what type of experiment is performed, these dynamics can become rather intricate.
In App.~\ref{app:levels} we consider two possible experimental methods that could be used for detecting the temperature-dependent qubit splitting and we show that Eq.~(\ref{Eq:DeltaE}) indeed captures the frequency-dependence of the extracted splitting very well.
We thus approximate the frequency near the resonance with this expression, expecting not more than small quantitative deviations very close to resonance.

The off-diagonal elements outside the red box in Eq.~\eqref{Eq:GeneralSingleModeHamiltonian}, coupling states $\ket{n-1,v_0}$ and $\ket{n,v_1}$, can also be included.
We treat them perturbatively, yielding corrections to the energy difference \begin{equation}
    \tilde{E}_n^{\rm oc} (g)= \frac{(2n+1)g^2}{\varepsilon+\hbar \omega}.
\end{equation}

As a next step, we consider a thermal distribution of phonons. Since bosonic operators commute for each~$\mathbf{k}$ in each branch~$\lambda$, the frequency can be calculated as in Eq.~\eqref{Eq:DeltaE} with $\tilde{E}_{\lambda,\mathbf{k},n}=	\tilde{E}_n|_{\omega=\omega_{\lambda}(\mathbf{k}),g=g_{\lambda}(\mathbf{k})}$ and $\tilde{E}_{\lambda,\mathbf{k},n}^{\rm oc}= \tilde{E}_n^{\rm oc}|_{\omega=\omega_{\lambda}(\mathbf{k}),g=g_{\lambda}(\mathbf{k})}$.
A frequency measured in an experiment will then be a statistic ensemble of frequencies corresponding to the Einstein-Bose statistics of the respective phonon frequency. 
The corrected energy splitting corresponding to the two levels $\ket{v_0}$ an $\ket{v_1}$ with a given phonon occupation number $n$ is then given by $\tilde{E}_{\lambda,\mathbf{k},n} + \tilde{E}_{\lambda,\mathbf{k},n}^{\rm oc}$.
To estimate the resulting frequency shift, we calculate the thermal average over all modes
\begin{align}
\langle \Delta \rangle 
= \sum_{\lambda,n} \frac{V}{(2 \pi)^3}\int {\rm d}^3k \, p_{\lambda,\mathbf{k},n} \left( \tilde{E}_{\lambda,\mathbf{k},n} + \tilde{E}_{\lambda,\mathbf{k},n}^{\rm oc} -\varepsilon \right). \label{Eq:GeneralPhononShift}
\end{align}
Here, we have introduced the probability of having $n$ phonons with frequency $\omega_{\lambda} (\mathbf{k})$ in branch $\lambda$ at temperature $T$ given by
\begin{align}
p_{\lambda,\mathbf{k},n} =  e^{-\frac{n \hbar \omega_{\lambda}(\mathbf{k})}{k_B T}} \left( 1-e^{-\frac{\hbar \omega_{\lambda}(\mathbf{k})}{k_B T}} \right), \label{Eq:PhononProbability}
\end{align}
where $k_B$ is the Boltzmann constant. We explicitly write $\langle \Delta \rangle$ to emphasize that we calculate a thermal average.

\section{Results}
\subsection{Spin-phonon coupling} \label{Sec:Spin-Phonon}
Since spin qubits typically operate at temperatures below $1\, {\rm K}$, we first focus on the smallest energy scale in a spin qubit, which is usually the Zeeman splitting $E_{z}$.  Let us assume that only low-energy phonons are present $\hbar \omega_{\lambda}(\mathbf{k}) \ll E_{v}$. In this case, since $|g_{\rm sv}|\ll E_{v}$ we can simplify the effective Hamiltonian in Eq.~\eqref{Eq:Heff-all} to 
\begin{align}
H = H_{z} + H_{p} + H_{\rm sp}, \label{Eq:Spin-Phonon-Hamiltonain}
\end{align}
consisting of the spin and phonon Hamiltonians, and an effective interaction between the two. Note that one can take first-order corrections due to the spin-valley coupling into account by substituting $g_{\rm sp} \rightarrow g_{\rm sp} - 2 g_{\rm sv} g_{\rm vp} /E_{v}$, where we have omitted the phonon indices. We find that Eq.~\eqref{Eq:Spin-Phonon-Hamiltonain} is equal to Eq.~\eqref{Eq:Hgeneral} with $\varepsilon = E_{z}$, $\ket{v_0}=\ket{\downarrow}$, $\ket{v_1} = \ket{\uparrow}$, $g=g_{\rm sp}$. According to Sec.~\ref{Sec:GeneralTreatment} the corrected eigenenergies corresponding to the spin-up and spin-down states with a given phonon occupation number~$n$ are then given by $\tilde{E}_{\lambda,\mathbf{k},n} (g_{{\rm sp}, \lambda, \mathbf{k}}) + \tilde{E}_{\lambda,\mathbf{k},n}^{\rm oc} (g_{{\rm sp}, \lambda, \mathbf{k}})$
and the resulting averaged qubit frequency shift is calculated by Eq.~\eqref{Eq:GeneralPhononShift}, $\langle \Delta_{\rm qubit} \rangle = \langle \Delta \rangle$. 

\begin{figure}
	\centering
	\includegraphics[width=0.47\textwidth]{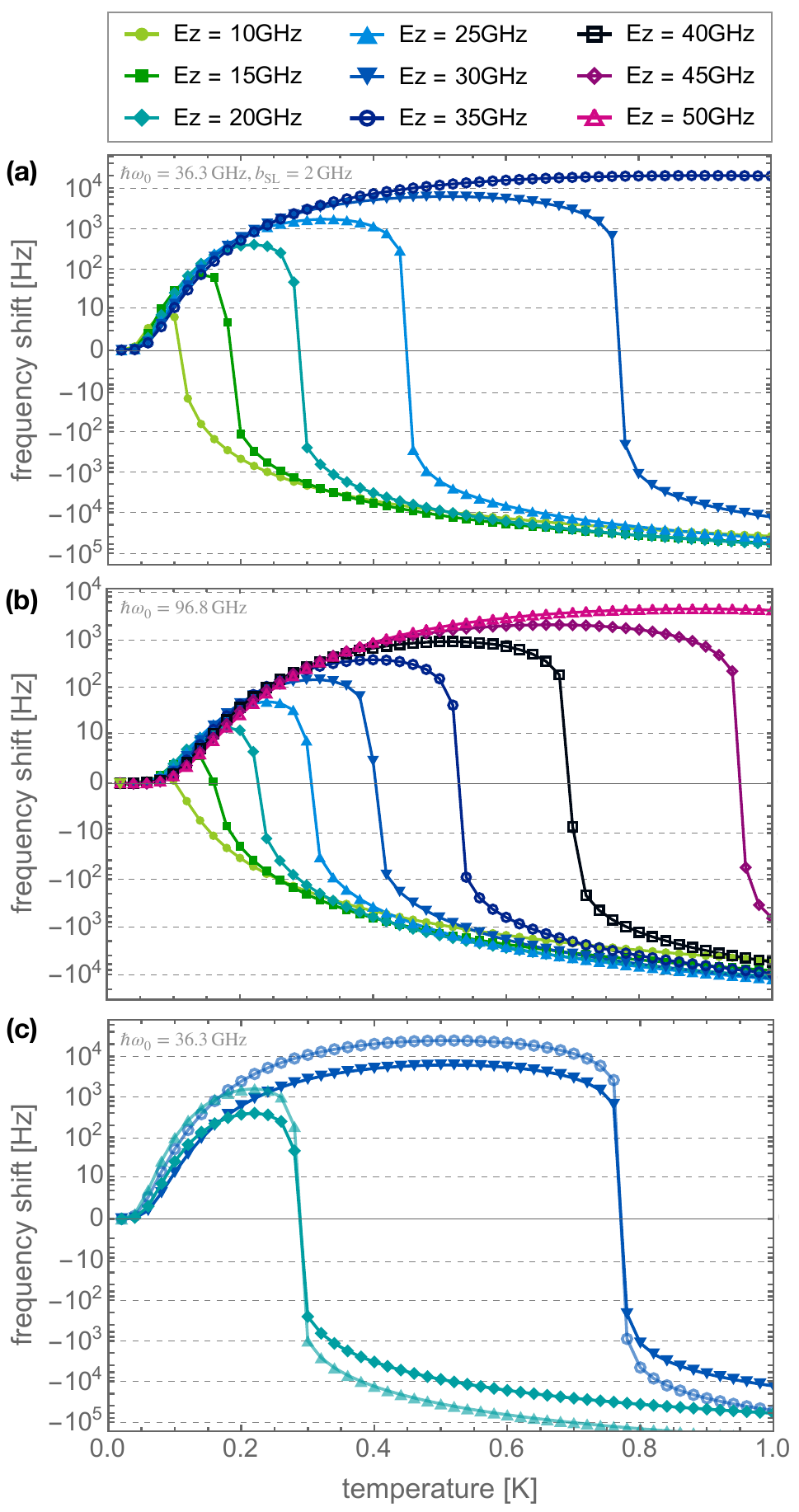}
	\caption{Temperature-dependent qubit energy shift calculated as in Eq.~\eqref{Eq:GeneralPhononShift} where we subtract the splitting at $T=0.02$~K. Shifts for various Zeeman fields $E_{z}$ at orbital splitting (a)~$\hbar \omega_{0,x}=0.15$~meV$~\approx 36.3$~GHz and $b_{\rm SL} = 2$~GHz, (b)~$\hbar \omega_{0,x}=0.4$~meV$~\approx 96.8$~GHz and $b_{\rm SL} = 2$~GHz. In (a) only results up to $E_{z} = 35$~GHz are shown as the approximation $E_{z} \ll \hbar \omega_{0,x}$ is violated with increasing $E_{z}$. (c)~Plots from (a) for $E_{z}=20$~GHz (turquoise) and 40~GHz (blue) with different values for $b_{\rm SL} = 2$~GHz and 4~GHz. For the $y$-axis scale we have used the scaling function ``SignedLog'' of Mathematica.}
	\label{Fig:PhononFreqShift}
\end{figure}

To estimate the frequency shift of the qubit and its dependence on the temperature, the integral over phonon energies is calculated numerically and the result is shown in Fig.~\ref{Fig:PhononFreqShift}. The cut-off for the number of phonons $n$ was taken to be 100~\footnote{The energies of phonons with $k_i \ell_i \sim 1$ are $\sim 0.1$--0.5~meV $\sim 1$--5~K, depending on the polarization of the phonon, for $\ell_i \approx 20$~nm.}, and $g_{\rm vo} =0$ was chosen for now \footnote{In our calculations we used $V=10^{-18}~{\rm m}^3$.
The actual value of $V$ should play no role, and we indeed found that our results are independent of the volume, as long as we set $V\lesssim 10^{-12}~{\rm m}^3$, above which numerical artifacts emerge.}. We show the temperature-dependent shift for varying Zeeman splitting $E_{z}$ for (a)~$\hbar \omega_{0,x}=0.15$~meV$~\approx 36.3$~GHz and $b_{\rm SL} = 2$~GHz, and (b)~$\hbar \omega_{0,x}=0.4$~meV$~\approx 96.8$~GHz and $b_{\rm SL} = 2$~GHz. We also show a critical dependence of the order of magnitude of the shift on $b_{\rm SL}$ in Fig.~\ref{Fig:PhononFreqShift}c. Here the frequency shifts for $b_{\rm SL} = 2$~GHz and 4~GHz are shown at two Zeeman fields $E_{z}=20$~GHz (turquoise) and 40~GHz (blue) at $\hbar \omega_{0,x}=0.15$~meV. 

At low temperatures, we find mostly positive contributions to the qubit frequency shift, as only low-energy phonons are occupied. With increasing temperature, more and more phonons with energy $\hbar \omega > E_z$ are occupied, which contribute with a negative shift and larger coupling $g$. Thus, we expect the qubit frequency shift to start decreasing and eventually even to switch sign, leading to a maximum in the curve, which our calculations show.
Furthermore, in our model the position of the maximum depends heavily on the Zeeman splitting. With increasing~$E_z$, the maximum occurs at higher temperatures as shown in Fig.~\ref{Fig:PhononFreqShift}. On the other hand, the maximum of the shift also depends on the orbital splitting since the latter defines the $\mathbf{k}$-dependence of the spin-phonon coupling, see Eq.~\eqref{Eq:Hsp}. Also, the order of magnitude of the shift depends on the size of the QD and thus the orbital splitting, but also other parameters such as $b_{\rm SL}$. As can be seen in Fig.~\ref{Fig:PhononFreqShift}c, when increasing $b_{\rm SL}$, the frequency shift increases. However, the maximum and sign change occur at the same temperature. We find the maximum of the curve to scale with $\propto b_{\rm SL}^2$. In Appendix~\ref{App:AnalyseData} we analyze the maximum frequency shift in more depth and show that the maximal shift scales as $\propto E_{z}^6$, $\omega_{0,x}^{-3}$, and $b_{\rm SL}^2$. 
Although the calculated shift is much smaller than the shifts obtained in experiments, we find a similar non-monotonic behavior with an initially increasing and then decreasing qubit frequency. 


Our theory assumes a thermal distribution of phonons. 
However, in experiments, temperature-dependent frequency shifts are often obtained using an off-resonant microwave burst employed prior to the measurement to change the temperature of the sample \cite{Takeda_2018, doi:10.1126/sciadv.add9408, Zwerver_2022, Freer_2017, PhysRevX.13.041015}. 
If a microwave pulse can increase the population of phonons at a specific energy determined by the driving frequency, the overall shift features more contributions from the respective phonons. Depending on whether the drive frequency lies above or below the Zeeman energy, we expect more negative or positive contributions to the shift, respectively.

We further note that unknown parameters, such as the exact value of $\omega_{0,x}$ and $b_{\rm SL}$, make a quantitative estimation of the shift difficult. We believe the approximations made are appropriate to estimate the effect due to phonons qualitatively. 
We have assumed the phonon dispersion of bulk silicon; however, in a heterostructure also  localized phonons at the quantum well interface can appear, which will affect the phonon density of states~\cite{Qi_2021, Cheng2021, Shi_2024}, and thus, the weight of specific shift contributions. This could give rise to an increase in the order of magnitude of the qubit frequency shift. 

The coupling $g_{\rm sp}$ can also be linked to the qubit relaxation time $T_1$ using Fermi's golden rule that is evaluated in Appendix~\ref{App:T1-time}.

\subsection{Valley-phonon coupling}
In the previous section, we have assumed that only low-energy phonons are occupied. In an ideal spin qubit, the valley splitting is rather large such that for many descriptions only the spin states may be considered. We now take into account phonons with slightly higher energies that might play a role with increasing temperature. To ensure that the Schrieffer-Wolff transformation in Sec.~\ref{Sec:Low-energy-supspace} is valid, we assume that the phonon energy is still small compared to the orbital splitting, $\omega_{\lambda} (\mathbf{k}) \ll \omega_{0,i}$.
In this regime, we investigate how the valley splitting is affected in the presence of phonons and what effect this will have on the spin energy splitting. In our model, we neglect that when phonons modulate the electron wavefunction in $z$-direction, this can result in an additional change of the valley splitting as the electron is pushed towards the heterostructure interface. Ultimately, this could also affect the spin qubit frequency.

\begin{figure}
	\centering
	\includegraphics[width=0.48\textwidth]{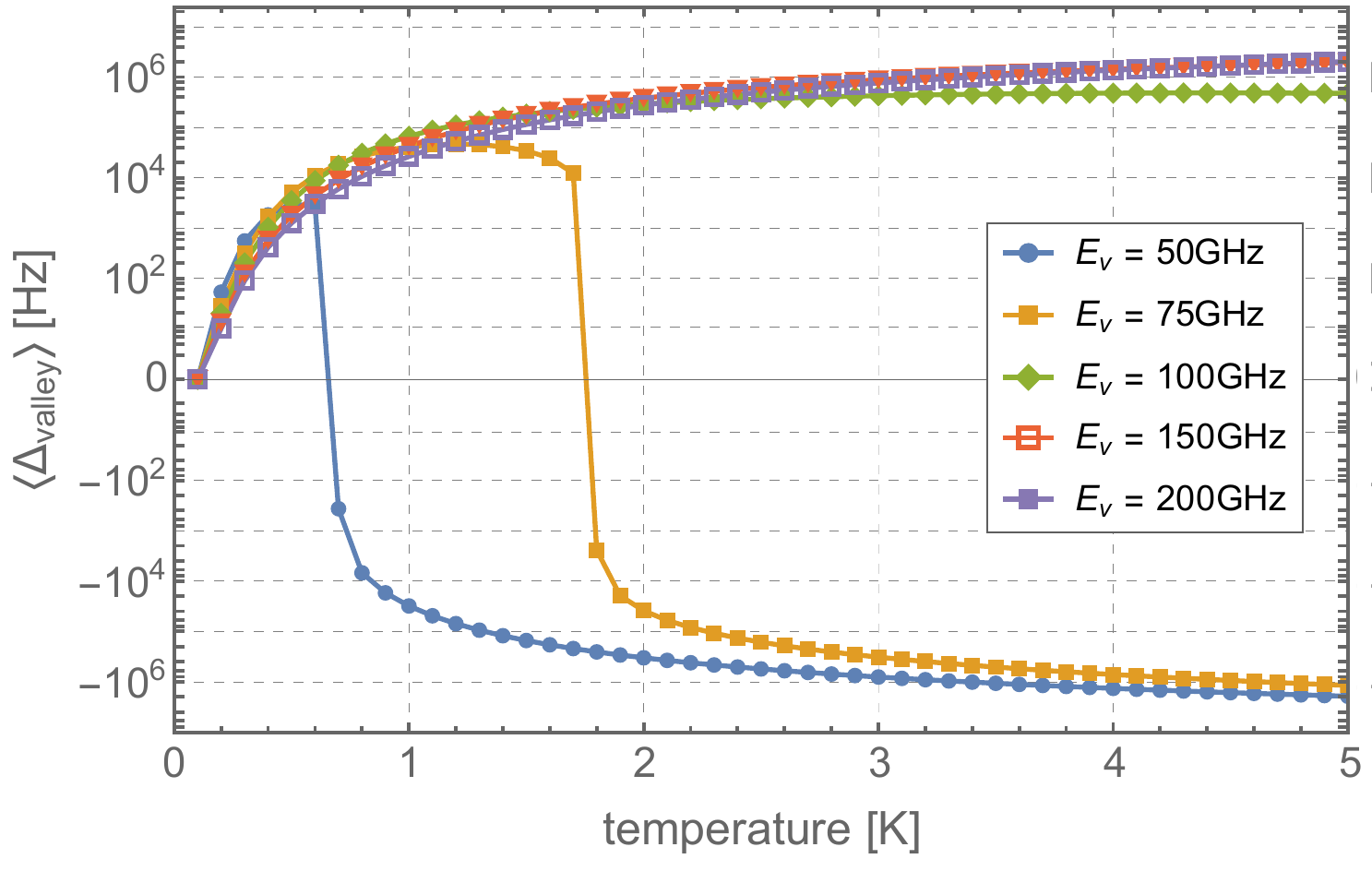}
	\caption{Temperature-dependent shift of the valley splitting where the shift at $T=0.1$~K is subtracted. We have used $\hbar \omega_{0,x}=1$~meV~$\approx 242$~GHz and $g_{\rm vo} = 10$~GHz. We find a non-monotonous shift where the maximum occurs at higher temperatures compared to the spin shift.  For the $y$-axis scale we have used the ``SignedLog'' scaling function of Mathematica.}
	\label{Fig:ValleyShift}
\end{figure}

First, we note that the smallest wave vector connecting the two low-lying valleys in the Brillouin zone is approximately $\mathbf{k} = 0.3 (2\pi/a_{\rm Si}) \mathbf{e}_z$. This corresponds to a phonon energy of $\hbar \omega_{l}(\mathbf{k}) \approx 32$~THz. For temperatures that are typical for spin qubit operations these phonons can be neglected. The energetically relevant phonons couple indirectly to the valley via the valley-orbit coupling.
For this we use the simplified Hamiltonian from Eq.~\eqref{Eq:Heff-all} and neglect the spin degree of freedom for now,
\begin{align}
    H= H_{v} + H_{p } + H_{\rm vp}.
\end{align}
Again we find that this corresponds to Eq.~\eqref{Eq:Hgeneral}, in this case with 
$\varepsilon = E_{v}$, $g=g_{\rm vp}$ and $\ket{v_0}=\ket{-z}$, $\ket{v_1} = \ket{+z}$ corresponding to the two low-lying valley states.
Following the procedure in Section~\ref{Sec:GeneralTreatment} the valley splitting is then shifted on average by $\langle \Delta_{\rm valley} \rangle = \langle \Delta \rangle$ as given in Eq.~\eqref{Eq:GeneralPhononShift}. We calculate the shift in the valley splitting in Fig~\ref{Fig:ValleyShift} for $\hbar \omega_{0,x}=1$~meV~$\approx 242$~GHz and $g_{\rm vo} = 10$~GHz. We find a similar behavior as in the previous section for the spin splitting but at higher temperatures. 

So far, we have neglected the spin degree of freedom. Obviously, the full effective Hamiltonian in Eq.~\eqref{Eq:Heff-all} contains a spin-valley coupling element $g_{\rm sv} \propto g_{\rm vo} b_{\rm SL}/\omega_{0,x}$. The direct spin-phonon coupling was evaluated in the previous section, and therefore we focus on the effective shift in the spin-qubit frequency due to the shift in valley splitting. For each phonon occupation $n$ in branch $\lambda$ with wave vector $\mathbf{k}$ we find a perturbative correction, 
\begin{align}
    \Delta_{\rm qubit, \lambda,\mathbf{k},n} = - \frac{2 g_{\rm sv}^2 E_{z}}{(\tilde{E}_{\lambda,\mathbf{k},n} + \tilde{E}_{\lambda,\mathbf{k},n}^{\rm oc})^2-E_{z}^2} + \frac{2 g_{\rm sv}^2 E_{z}}{E_{v}^2-E_{z}^2}, \label{Eq:Valley-Qubit-Shift-per-Mode}
\end{align}
to the qubit energy. 
From Fig.~\ref{Fig:ValleyShift} we can estimate the resulting spin shift due to the spin-valley coupling. In the regime of the approximations made throughout the paper, the deviations are on the order of up to $10^{2}$ Hz. However, at the spin-valley hotspot where $E_z\approx E_v$, the qubit frequency shift could range up to higher values. At the same time, we expect the approximations to break down in this regime, such that higher-order corrections must be taken into account or the combined spin-valley subsystem needs to be solved.

\subsection{Electron-phonon coupling} \label{Sec:Electron-Phonon}
Finally, we aim to understand the role of the orbital splitting for the spin qubit frequency. With increasing temperature, more and more phonons with higher energy are present in the quantum well of the heterostructure. In this case, we cannot rely on the Schrieffer-Wolff transformation applied for the previous description, and start with Fig.~\ref{Fig:Model}a. In the following, we neglect the resonance of low-energy phonons around the qubit energy and the valley splitting captured in previous sections, and focus on phonon energies of the order of the orbital splitting $\hbar \omega \lessapprox \hbar \omega_{0,x}$. For this, we neglect both the Zeeman splitting and the valley degree of freedom for now and describe the system using the Hamiltonian,
\begin{align}
H = H_{\rm QD} + H_{p} + H_{\rm ep}, \label{Eq:OrbitalHamiltonian}
\end{align}
consisting of the quantum dot confinement potential, the phonon Hamiltonian, and the electron-phonon coupling. As we are only interested in the lowest-lying orbital states we cut the Hamiltonian after the first excited states and leave out $n_{x}, n_{y}, n_{z} >1$. With that, we neglect two-phonon processes as well as higher-energy phonons entirely and use Eq.~\eqref{Eq:Hep-1stexcitedstate} for $H_{\rm ep}$.
Exemplarily, we can write the $x$-components in Eq.~\eqref{Eq:OrbitalHamiltonian} as a two-level system interacting with phonons as in Eq.~\eqref{Eq:Hgeneral} with $\varepsilon = \omega_{0,x}/2$, $\ket{v_0}=\ket{F_{000}}$, $\ket{v_1} = \ket{F_{100}}$, and $g=C_{x}$, where we again omitted the phonon indices.

In the limit of $|\sqrt{n} C_x| \ll \hbar |\omega_{0,x}+\omega|$, which holds for the temperatures in spin qubit devices, we can again follow the procedure outlined in Sec.~\ref{Sec:GeneralTreatment}. The resulting orbital frequency is then given by $\langle \Delta_{\rm orbit} \rangle = \langle \Delta \rangle$.

Similarly to the previous section, the shift in energy does not appear as a direct shift in the qubit frequency, but as a shift in the orbital splitting of the quantum dot. Using perturbation theory, this corrects the qubit frequency analogously to Eq.~\eqref{Eq:Valley-Qubit-Shift-per-Mode} by
\begin{align}
    \Delta_{\rm qubit, \lambda,\mathbf{k},n} = - \frac{2 b_{\rm SL}^2 E_{z}}{(\tilde{E}_{\lambda,\mathbf{k},n} + \tilde{E}_{\lambda,\mathbf{k},n}^{\rm oc})^2-E_{z}^2} + \frac{2 b_{\rm SL}^2 E_{z}}{\hbar^2 \omega_{0,x}^2-E_{z}^2}, \label{Eq:Orbital-Qubit-Shift-per-Mode}
\end{align}
due to the spin-orbit coupling element $b_{\rm SL}$. The thermal average will then result in a temperature dependence.

Since the expression in Eq.~\eqref{Eq:OrbitalHamiltonian} is analogous to Eq.~\eqref{Eq:Hgeneral}, low-energy phonons ($\omega < \omega_0$) increase the orbital shift and high energy phonons ($\omega > \omega_0$) decrease the orbital shift. In equation~\eqref{Eq:Orbital-Qubit-Shift-per-Mode}, the orbital splitting adds with a negative sign and inversely proportional to the qubit frequency shift. As a result, phonon energies with $\omega < \omega_0$ induce a positive qubit frequency shift, and phonon energies with $\omega > \omega_0$ a negative shift. We expect a similar behavior as in previous section on larger temperature range. In the temperature range of interest for operating spin qubits, this effect will presumably be small.

\section{Summary}
In this paper, we have investigated the effect of phonons on the energy splitting of an electron spin qubit confined in a Si quantum dot. We have considered different energy scales on which phonons can interact with the system.
We started with a low-temperature model that effectively describes the spin-phonon coupling. 
This we have derived from the interaction of the electron with the deformation potential of the phonons and the artificially induced spin-orbit coupling due to a magnetic gradient across the device, and we also found corrections in the effective coupling due to the presence of valley-orbit coupling. 
With our model, we find that each phonon mode contributes to an overall frequency shift of the qubit. Phonons with energies lower than the Zeeman splitting lead to a positive shift, whereas phonons with higher energies lead to a negative shift. We thus find a non-monotonic behavior, similar to the experimentally observed behavior. To obtain explicit results, we have calculated the effective coupling and the estimated frequency shift numerically. We find intriguing dependencies of the shift on various parameters, such as the Zeeman splitting and the quantum dot size, $\langle \Delta_{\rm qubit}\rangle \propto E_{z}^6$, $\omega_{0,x}^{-3}$, and $b_{\rm SL}^2$.

Furthermore, we have considered the valley and orbital splittings as the second and third energy scales on which phonons can indirectly interact with the qubit system. The resulting shift of the orbital level ultimately yields a shift in the qubit frequency. We find the valley and orbital splitting to determine the size and temperature behavior of the respective shifts. However, since this mechanism requires rather high-energy phonons, we expect this effect to appear at higher temperatures compared to the effective spin-phonon coupling case, as our calculations indicate. 

In conclusion, our findings indicate that phonons can have a temperature-dependent impact on the qubit frequency. However, with realistic parameters that are typical for spin qubits, we do not expect the effect to be on the same order of magnitude as in many experiments
such as Refs.~\cite{PhysRevApplied.19.044078, PhysRevX.13.041015, champain2025heatresilientholespinqubit, ye2025measuringpulseheatingsi}. One way to test our theory is to measure a temperature-dependent frequency shift at different magnetic fields and dot sizes. 


As the exact origin of temperature-dependent frequency shifts in spin-qubit devices is unknown, there are other possible explanations, such as electric dipoles~\cite{champain2025heatresilientholespinqubit, PhysRevResearch.6.013168} due to charge traps at the material interface. The thermal activation of these two-level fluctuators can have a similar effect on the qubit frequency shift with an analog temperature dependence as presented in this work. A non-monotonous shift could also be explained via electric dipoles with corresponding energies below and above the spin splitting, following the arguments in our work and featuring a similar magnetic-field dependence that should be further investigated in future experiments.

\section*{Acknowledgments}
We would like to thank 
Irene Fernandez De Fuentes, 
Daniel Loss,
John Morton,
Jason R. Petta,
Maximilian Rimbach-Russ,
Brennan Undseth, 
Florian Unseld, and
Lieven Vandersypen for the insightful discussions on various aspects of this project. 
This work was supported by the Army Research office (ARO) under grant W911NF-23-1-0104.

\appendix

\section{Spin relaxation rate} \label{App:T1-time}
The coupling $g_{\rm sp}$ can be linked to the qubit relaxation time $T_1$ using Fermi's golden rule. Therefore, we consider an initial product state $\ket{\lambda,\mathbf{k},n \uparrow }$ with spin-up and $n$ phonons with wave vector $\mathbf{k}$ in branch $\lambda$ with eigenenergy $E_{\lambda,\mathbf{k},n,\uparrow}$. The probability for the respective phonon state is given by $p_{\lambda,\mathbf{k},n}$. 
We calculate the relaxation rate 
\begin{widetext}
\begin{align}
\frac{1}{T_1} &= 
\frac{2\pi}{\hbar} \sum_{\lambda',\mathbf{q},m} \sum_{\lambda,\mathbf{k},n} p_{\lambda,\mathbf{k},n}  |  \bra{\lambda',\mathbf{q},m \downarrow } H_{\rm int} \ket{\lambda,\mathbf{k},n \uparrow }|^2 \delta (E_{\lambda,\mathbf{k},n,\uparrow} - E_{\lambda',\mathbf{q},m,\downarrow}) \label{Eq:FermisGoldenRule2}\\
&= \frac{ V}{4 \pi^2 \hbar^2} \sum_{\lambda,\mathbf{k} }  \int k^2{\rm d}k \int \sin(\theta) {\rm d}\theta {\rm d} \phi  \frac{\left| g_{\lambda} \left( \mathbf{k} \right)\right|^2}{v_{\lambda}} \left( \langle n (\hbar v_\lambda k ) \rangle +1 \right) \delta \left( k- \frac{E_{z}}{\hbar v_{\lambda}} \right) \\
&= \frac{E_{z}^5 b_{\rm SL}^2 \ell^2}{4\pi^2 \hbar ^8 \omega_0^2 \rho_{\rm Si}} \sum_{\lambda=l,t} \frac{I_{\lambda}}{v_{\lambda}^7} \left(\langle n (E_{z}) \rangle +1 \right) e^{-\frac{\ell^2E_{z}^2}{2 \hbar^2 v_{\lambda}^2}}, \label{App:Eq:g-estimation}
\end{align}
\end{widetext} 
where we have assumed a linear dispersion $\omega_{\lambda}(k) = v_{\lambda} k$. Here, $\langle n (E_{z}) \rangle = (e^{E_{z}/(k_B T)} -1)^{-1}$ is the average phonon number with energy $E_{z}$ at temperature $T$ and 
\begin{align}
	I_l &= \int_{0}^{2\pi} \int_{0}^{\pi} (\Xi_{d} + \Xi_{u} \cos^2(\theta))^2  \sin^3(\theta) \cos^2(\phi) {\rm d}\theta {\rm d} \phi \notag \\
	&= \frac{4\pi}{105} \left( 35 \Xi_d^2 +14 \Xi_d \Xi_u + 3\Xi_u^2 \right),\\
	I_t &= \int_{0}^{2\pi} \int_{0}^{\pi} ( \Xi_{u} \sin(\theta) \cos(\theta))^2  \sin^3(\theta) \cos^2(\phi) {\rm d}\theta {\rm d} \phi \notag \\
	&= \frac{16 \pi \Xi_d^2}{105},
\end{align}
are the surface integrals, where we have assumed an isotropic dispersion relation.

\section{Electron-phonon coupling and validity of the approximations} \label{App:PhononsModelValidity}

\begin{figure}
	\centering
	\includegraphics[width=0.49\textwidth]{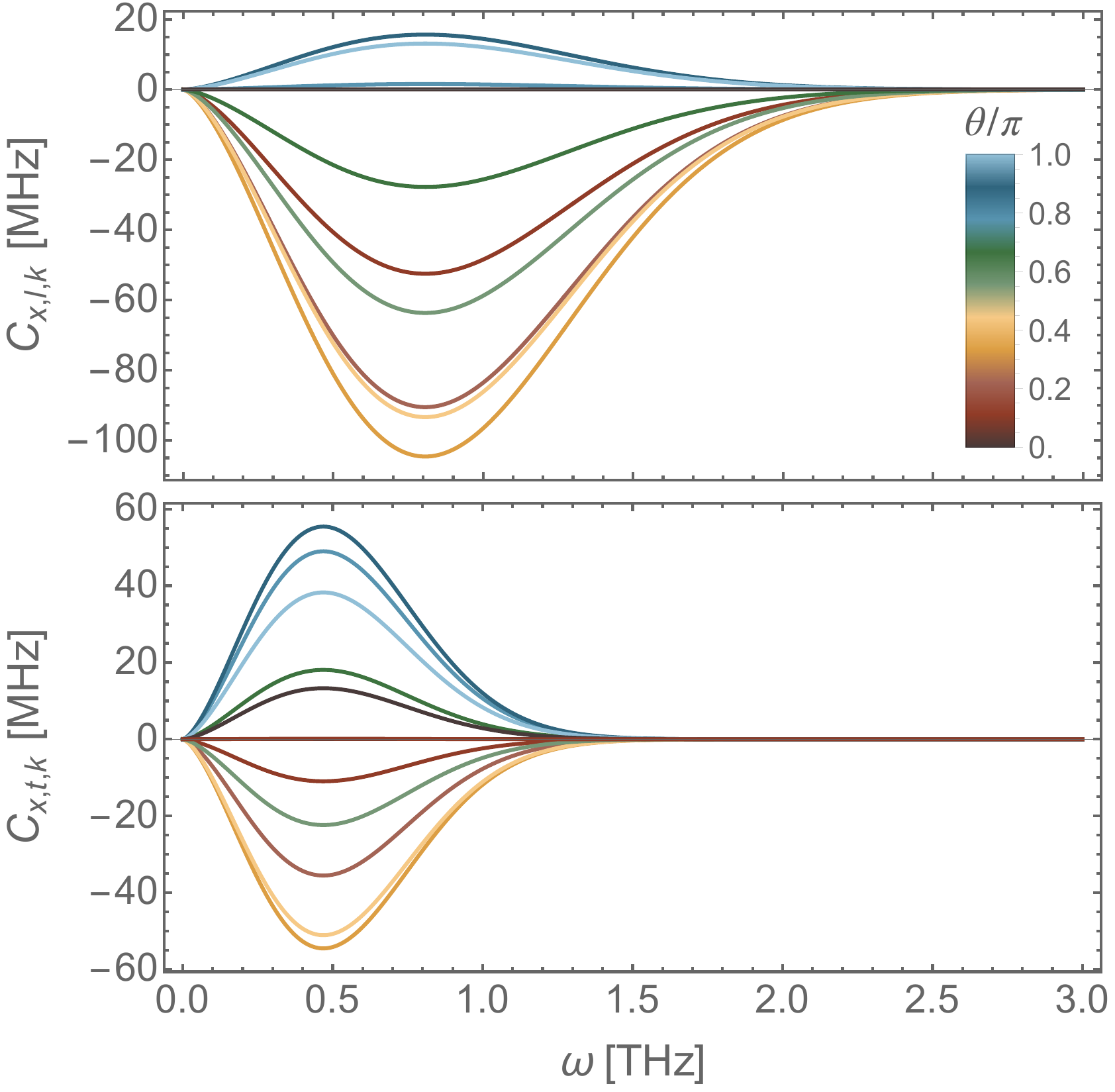}
	\caption{Longitudinal and transversal electron-phonon couplings $C_{x,l,\bf{k}}$ and $C_{x,t,\bf{k}}$ depending on the phonon energy $\hbar \omega=\hbar v_{l,t} k$ with orbital splitting $\hbar \omega_0 = 1$~meV $\approx 242$~GHz and volume $V=10^{-18}$~$\rm m^3$. For low energy phonons, $\omega \ll \omega_0$, the coupling shows the $\omega^{3/2}$ behavior one obtains in an electric-dipole approximation, whereas for high-energy phonons, the wave vector is too large to couple to the quantum dot orbitals. Note that we only consider excitations to the first excited orbital state as spin qubits operate at rather low temperatures. Here, we have used $\phi=0$ for simplicity, such that $\mathbf{k}=k(\sin(\theta), 0, \cos(\theta))$.}
	\label{Fig:PhononCoupling}
\end{figure}

To calculate the shift in qubit frequency, we have used assumptions that we elaborate here in more detail. In silicon spin qubits, the Zeeman splitting is usually the smallest energy scale, followed by the valley splitting, which is usually lower than the orbital splitting $E_{z} \ll E_{v} \ll \omega_{0,i}$. For small couplings to the first orbital $|g_{\rm vo}|, |b_{\rm SL}|, |C_{i,\lambda,\bf{k}}| \ll \omega_{0,i}$ the Schrieffer-Wolff transformation is justified. The valley-orbit coupling arises due to imperfections at the heterostructure interface and can be on the order of up to $\sim 10^{10}$~Hz. The micromagnet-induced spin-orbit coupling due to a slanting magnetic field is typically $\sim 10^8$~Hz. The electron-phonon coupling in the $x$-direction is shown in Fig.~\ref{Fig:PhononCoupling} for the transversal and longitudinal phonon branches in silicon. It depends on the phonon energy $\omega_{\lambda}(\mathbf{k}) = v_{\lambda}k$ where the wave vector is $\mathbf{k}=k(\sin(\theta) \cos(\phi), \sin(\theta)\sin(\phi), \cos(\theta))$, and thus changes with $\theta$ and $\phi$. For simplicity, we have chosen $\phi = 0$ in the plot. In the low-energy regime the electric-dipole approximation holds and exhibits a $\propto k^{3/2}$ behavior. If the phonon wave vector is much larger than the size of the quantum dot the coupling decreases to zero. As spin qubits are operated at temperatures around hundreds of millikelvins, two-phonon processes and excitations to higher orbital states are neglected in our calculations. 

In the approximations made in this paper, we have to ensure that $\sqrt{n} g_{\lambda, \mathbf{k}} \ll \varepsilon$. In particular, we need to justify $\sqrt{n} 2 b_{\rm SL} C_{x, \lambda, \mathbf{k}} / \omega_{0,x} \ll E_{z}$, $\sqrt{n} g_{\rm vo} C_{x, \lambda, \mathbf{k}}/\omega_{0,x} \ll E_{v}$ and $\sqrt{n} C_{x, \lambda, \mathbf{k}} \ll \hbar \omega_0$. 

\begin{figure}[b]
	\centering
	\includegraphics[width=0.48\textwidth]{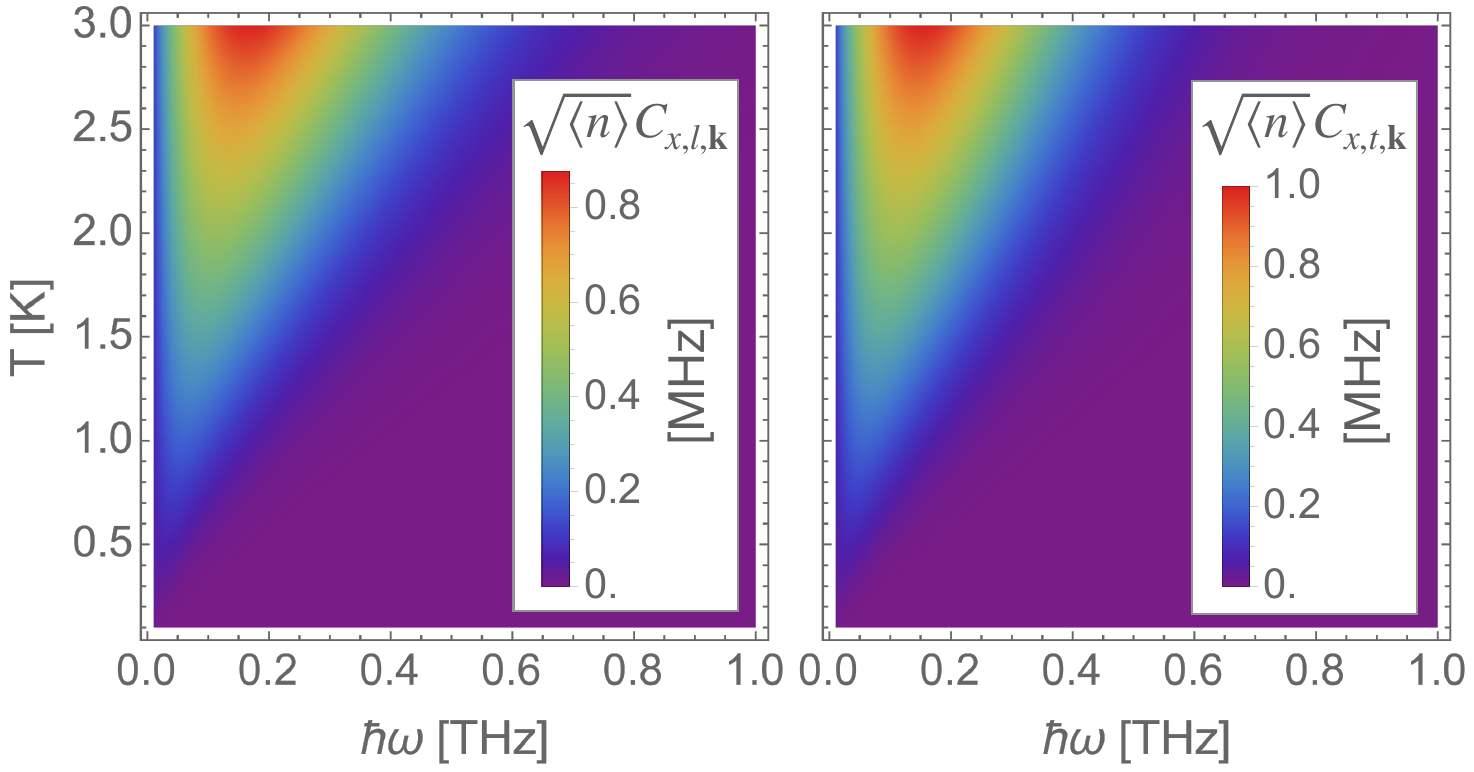}
	\caption{Numerical value of $\sqrt{\langle n \rangle} C_{x,l,\mathbf{k}}$ and $\sqrt{\langle n \rangle} C_{x,t,\mathbf{k}}$ depending on the phonon energy $\hbar \omega$ and temperature $T$, where we have chosen $V=10^{-18}$~m$^3$, $\hbar \omega_0=1$~meV$\approx 242$~GHz, $\phi = 0$ and we have maximized the coupling with respect to $\theta$.}
	\label{Fig:PhononApproximationValidation}
\end{figure}

To show that our model is valid in the temperature regime of interest, we calculate $\sqrt{\langle n \rangle} C_{x,\lambda,\mathbf{k}}$, where $\langle n \rangle = (e^{v_{\lambda}k/(k_B T)} -1)^{-1}$ is the average phonon occupation number, for the transverse coupling explicitly, and plot it as a function of the phonon energy $\hbar \omega$ and temperature $T$ in Fig.~\ref{Fig:PhononApproximationValidation}. 
We find $\sqrt{\langle n \rangle} C_{x,\lambda,\mathbf{k}}/\omega_0 \leq 10^{-5} \ll 1 $, and with $b_{\rm SL}/E_{z} < 1$, and $g_{\rm vo}/ E_{v} <1$ each of the stated approximations is valid.

\section{Analysis of maximal frequency shift} \label{App:AnalyseData}
In addition to Fig.~\ref{Fig:PhononFreqShift}, we analyze dependencies of the calculated frequency shift on the Zeeman field $E_{z}$, the orbital splitting $\omega_{0,x}$ and the slanting magnetic field responsible for spin-obit coupling. First, we consider the spin-phonon coupling $g_{\rm sp} (\mathbf{k}) \propto b_{\rm SL} k^{3/2} e^{- \ell_x^2 k^2} / \omega_{0,x}^{3/2}$. For off resonant phonons we can approximate the qubit shift $\Delta_{\rm qubit} \propto g^2$ as in Eq.~\eqref{Eq:Lamb-shift}, and thus $\Delta_{\rm qubit} \propto b_{\rm SL}^2 k^{3} e^{- \hbar k^2/(2 m_t\omega_{0,x})} / \omega_{0,x}^{3}$. We will use these relations to estimate the maximal shift in the following.

\begin{figure*}
	\centering
	\includegraphics[width=0.9\textwidth]{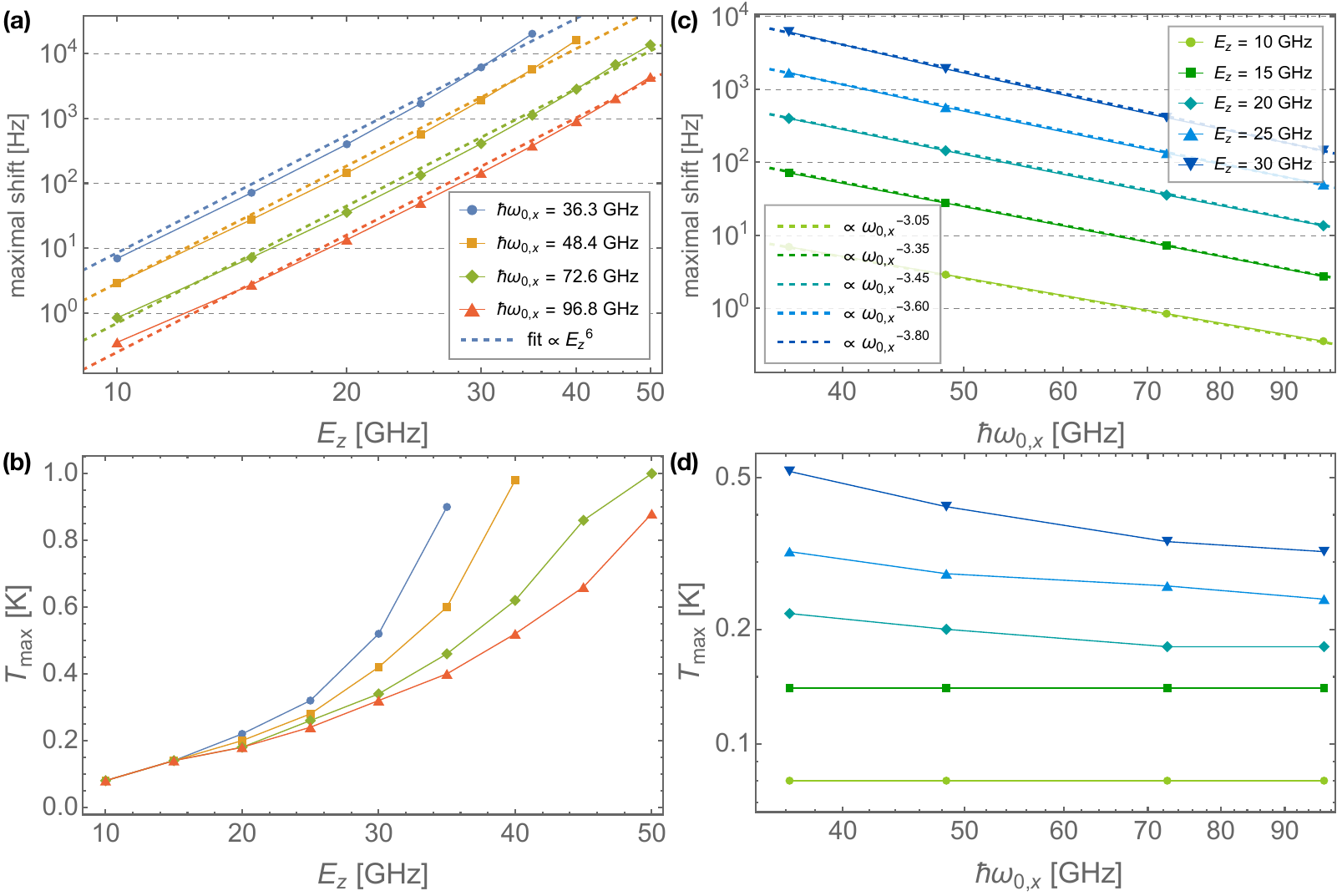}
	\caption{Extended analysis of the maximum in Fig.~\ref{Fig:PhononFreqShift}. (a) Maximal frequency shift and (b) position of the maximum frequency shift depending on the Zeeman field $E_{z}$ for varying $\omega_{0,x}$. (c) Maximal frequency shift and (d) position of the maximum frequency shift depending on the orbital splitting $\omega_{0,x}$ for varying $E_{z}$. Data points correspond to numerical calculations and dashed lines to fits as discussed in Appendix~\ref{App:AnalyseData}. All calculations assume $b_{\rm SL} = 2$~GHz.}
	\label{Fig:Analysis1}
\end{figure*}

In Fig.~\ref{Fig:Analysis1} we plot (a)~the maximal frequency shift and (b)~the position of the maximum frequency shift depending on the Zeeman field $E_{z}$ for varying $\omega_{0,x}$. In the regime $E_{z} =10 - 50$~GHz, for the maximal frequency shift, we can approximately fit a $\propto E_{z}^6$ dependency. We note, however, that already in this small range we find deviations from this fit. At small phonon frequenies, $\omega \ll \omega_{0,x}$ we can assume $g_{\rm sp} \propto \omega^{3/2}$. For off resonant phonons, the shift scales with $g_{\rm sp}^2 \propto \omega^{3}$. Integration yields another $\omega^3$. As $E_{z}$ defines up to which phonons contribute positively to the shift, we can approximate $\omega \sim E_{z}$. We expect a further deviation from this behavior as the Zeeman field $E_{z}$ approaches the orbital splitting $\hbar \omega_{0,x}$. The temperature~$T_{\rm max}$ at which the maximum of the shift occurs shows an increase with increasing Zeeman field, but could not be fit to a simple relation.

In Fig.~\ref{Fig:Analysis1} we show (c)~the maximal frequency shift and (d)~the position of the maximum frequency shift depending on the orbital splitting $\omega_{0,x}$ for various $E_{z}$. For the maximum value of the shift, we find slightly different power laws around $\omega_{0,x}^{-3}$ for different $E_{z}$ that we assume to change more drastically with $E_{z}$ approaching $\hbar \omega_{0,x}$. The power law of approximately $\omega_{0,x}^{-3}$ fits well with our expectations discussed above. In Fig.~\ref{Fig:Analysis1}d we find that for increasing orbital splitting the maximum shifts towards lower temperatures~$T_{\rm max}$. Again, we find a change in the curves as $E_{z}$ approached $\hbar \omega_{0,x}$.
All calculations discussed above assume $b_{\rm SL} = 2$~GHz. 

\begin{figure}
	\centering
	\includegraphics[width=0.48\textwidth]{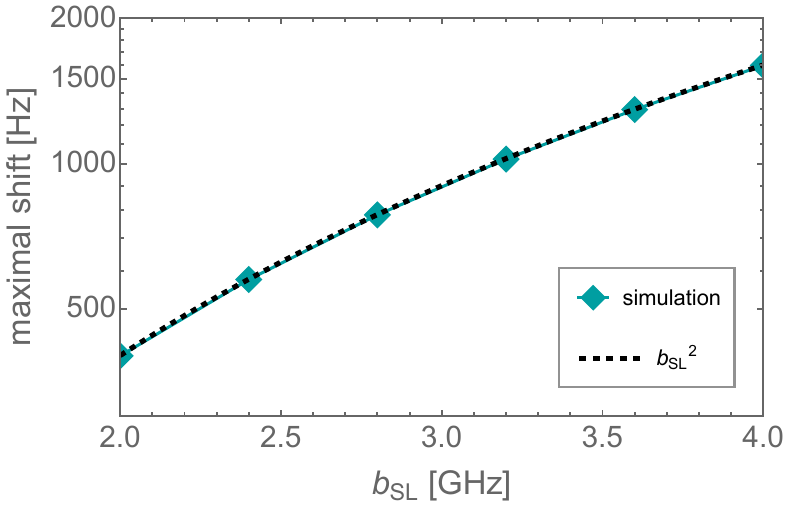}
	\caption{Maximal frequency shift depending on $b_{\rm SL}$ where $\hbar \omega_{0,x} = 36.6$~GHz and $E_{z} = 20$~GHz. The simulated data points perfectly match the expected $b_{\rm SL}^2$ behavior.}
	\label{Fig:Analysis2}
\end{figure}

Next, we identify the relation between the maximal shift and the slanting field, which introduces spin-orbit coupling. Different from the previous parameters, $b_{\rm SL}$ only appears linearly in the phonon coupling and is independent of the phonon occupation distribution. As Fig.~\ref{Fig:PhononFreqShift}b shows, the position of the maximum is independent of $b_{\rm SL}$. However, the order of magnitude of the maximal shift depends quadratically on $b_{\rm SL}$. Figure~\ref{Fig:Analysis2} shows simulation points of the maximal shift value for different~$b_{\rm SL}$ and features a perfect match with the quadratic fit function shown as black dashed lines.

\section{Experimental extraction of effective qubit splitting}\label{app:levels}

\begin{figure}[t!]
	\centering
	\includegraphics[width=0.48\textwidth]{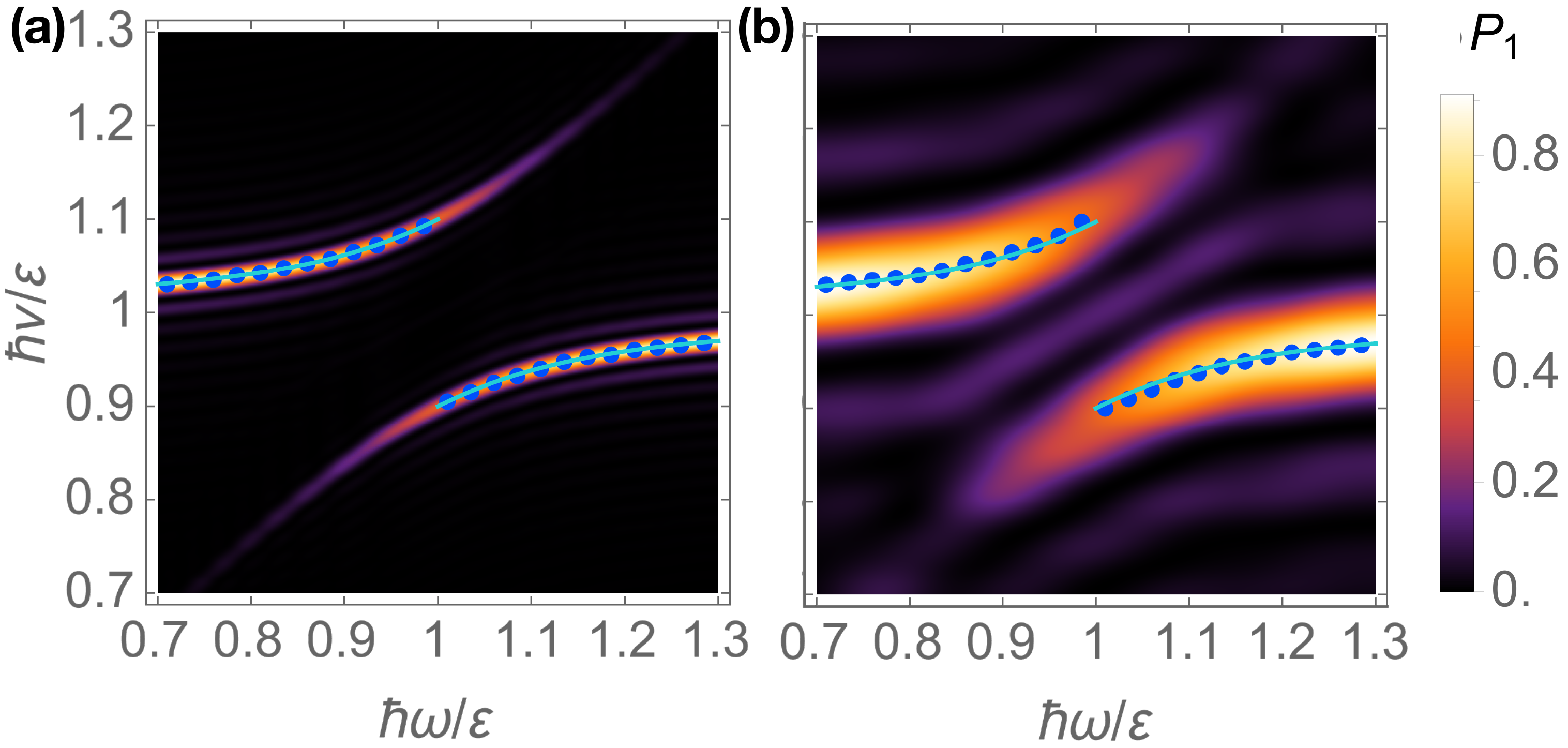}
	\caption{Numerically calculated probability $P_1$ to find a qubit initialized in the ground state $\ket{0,v_0}$ in its excited state after a Rabi $\pi$-pulse, as a function of driving frequency $\nu$ and phonon frequency $\omega$.
    We used $g = 0.1\,\varepsilon$ and driving strength (a)~$A/\varepsilon = 0.02$ and (b)~$A/\varepsilon = 0.1$.
    The blue dots represent the qubit splitting extracted as the maximum $P_1$ for each $\omega$.
    The solid curves show the qubit splitting as described by Eq.~(\ref{Eq:DeltaE}).}
	\label{Fig:Ramsey}
\end{figure}

In this Appendix we connect the frequency-dependent qubit splitting given in Eq.~(\ref{Eq:DeltaE}) to experimentally observable quantities.
The dynamics described by the Hamiltonian (\ref{Eq:Hgeneral}) can be rather complex close to the resonance condition and the choice to simply define the effective qubit splitting to be the energy difference between the eigenstates with the strongest components $\ket{n,v_0}$ and $\ket{n,v_1}$ might seem a bit arbitrary at first sight.

Indeed, since at resonance all eigenstates will contain equal weights of both qubit states $\ket{v_0}$ and $\ket{v_1}$, it becomes strictly speaking impossible to define a ``qubit splitting'' when $\hbar\omega$ approaches $\varepsilon$.
It is thus instructive to consider possible experimental procedures for extracting the qubit splitting and ask the question how a system described by the model Hamiltonian (\ref{Eq:Hgeneral}) would respond to each procedure.

For simplicity, we will focus on the qubit splitting for the case of $n=0$, where the state $\ket{1,v_0}$ anticrosses with the ``bare'' excited qubit state $\ket{0,v_1}$ at $\hbar\omega = \varepsilon$.
One method for determining a qubit splitting is to perform a series of Ramsey experiments, where the qubit is initialized in its ground state, then subjected to a $\pi/2$-pulse along $\hat \sigma_y$, let evolve freely for a time $\tau$, subjected to another $\pi/2$-pulse along $\hat \sigma_y$, and then measured.
The $\tau$-dependent probability $P_0(\tau)$ to measure the qubit in its ground state is expected to be $\sin^2 (\varepsilon \tau/2\hbar)$ and a series of such Ramsey experiments for different $\tau$ will thus allow to determine the qubit splitting $\varepsilon$ by extracting the frequency of the oscillations observed in $P_0(\tau)$.

We consider such a series of experiments, restricting ourselves to the subspace $\{\ket{0,v_0},\ket{0,v_1},\ket{1,v_0},\ket{1,v_1}\}$.
We initialize the system in $\ket{0,v_0}$ at $t=0$, apply the rotation $e^{-i \frac{\pi}{4} \hat\sigma_y}$, let the system evolve for time $\tau$ under the Hamiltonian
\begin{equation}
    \hat H_{4\times 4} = \left( \begin{array}{cccc}
    0 & 0 & 0 & 0 \\
    0 & \varepsilon & g & 0 \\
    0 & g & \hbar\omega & 0 \\
    0 & 0 & 0 & \varepsilon + \hbar\omega
    \end{array}\right),\label{eq:h4x4}
\end{equation}
apply another rotation $e^{-i \frac{\pi}{4} \hat\sigma_y}$, and calculate the total probability to find the qubit in the ground state $P_0(\tau) = |\bra{\psi(t)}\ket{1,v_0}|^2 + |\bra{\psi(t)}\ket{0,v_0}|^2$.
We can find explicitly
\begin{align}
    P_0(\tau) = {} & {} 
    \frac{1}{2}\bigg[ 1 - \cos(\tfrac{1}{2}\alpha\tau)\cos(\tfrac{1}{2}[\hbar\omega+ \varepsilon]\tau) \nonumber\\ {} & {} 
    \hspace{1em}+ \frac{\varepsilon-\hbar\omega}{\alpha} \sin(\tfrac{1}{2}\alpha\tau)\sin(\tfrac{1}{2}[\hbar\omega+ \varepsilon]\tau)\bigg] \nonumber\\ {} & {}
    + \frac{15 g^2}{16 \alpha^2}\sin\left(\frac{\alpha \tau}{2}\right)^2,
\end{align}
with $\alpha = \sqrt{4g^2 + (\varepsilon-\hbar\omega)^2}$.
It is straightforward to take the Fourier transform with respect to $\tau$ of this expression, which consists of a sum of Dirac delta functions, and it is then easy to see that the dominating component is indeed at the frequency $hf = \frac{1}{2}(\varepsilon + \hbar\omega) \pm \frac{1}{2}\sqrt{4g^2 + (\varepsilon-\hbar\omega)^2}$, where the $+$ ($-$) sign applies to the region $\hbar\omega < \varepsilon$ ($\hbar\omega > \varepsilon$), exactly as described by Eq.~(\ref{Eq:DeltaE}).


An alternative experimental method for determining the qubit splitting could be to probe its response to Rabi driving.
One could initialize the qubit in the ground state and then apply a microwave pulse $\frac{1}{2} A \cos(\nu t)\hat\sigma_x$ of length $T = h / A$.
This pulse is expected to implement a $\pi$-rotation of the qubit along $\hat\sigma_x$ when the driving frequency is at resonance with the qubit splitting.
The frequency for which the probability to end up in the excited qubit state is found to be largest can then be identified as the qubit frequency.

We also simulate this type of experiment, again focusing on the subspace $\{\ket{0,v_0},\ket{0,v_1},\ket{1,v_0},\ket{1,v_1}\}$ for simplicity.
We initialize the system in $\ket{0,v_0}$ at $t=0$, let the system evolve under the Hamiltonian $\hat H_{4\times 4} + \frac{1}{2} A \cos(\nu t)\hat\sigma_x$ for time $T = h/A$, where $\hat H_{4\times 4}$ is given in Eq.~\eqref{eq:h4x4}, and we set $g=0.1\,\varepsilon$ again.
In Fig.~\ref{Fig:Ramsey}a,b we plot the final probability to find the qubit in the excited state, $P_1(h/A) = |\bra{\psi(h/A)}\ket{1,v_1}|^2 + |\bra{\psi(h/A)}\ket{0,v_1}|^2$, as a function of $\hbar\nu$ and $\hbar\omega$ (color plots).
The blue dots indicate the numerically extracted location of the maximal $P_1$ as a function of frequency $\nu$ for each $\omega$; the qubit splitting as given by Eq.~(\ref{Eq:DeltaE}) is plotted in solid blue.
We show results for two different driving strengths, (a)~$A/\varepsilon = 0.02$ and (b)~$A/\varepsilon = 0.1$.
We see that also in this case the qubit splitting that would be extracted around the anticrossing matches Eq.~(\ref{Eq:DeltaE}) very well.

\bibliography{bibliography}

\end{document}